\documentclass[onecolumn,showkeys, prb]{revtex4}
\usepackage{amsfonts, amsmath, latexsym}
\usepackage[dvips]{graphicx}

\newtheorem{1d}{Definition}
\newtheorem{1}{Theorem}
\newtheorem{2d}[1d]{Definition}
\newtheorem{1p}{Proposition}

\newtheorem{3p}[1p]{Proposition}
\newtheorem{2}[1]{Theorem}
\newtheorem{3}[1]{Theorem}
\newtheorem{1c}{Corollary}

\begin{document}

\title{Generalized moments of spectral functions from short-time correlation functions}

\author{Cristian Predescu}
\email{cpredescu@comcast.net} 
\affiliation{Department of Chemistry and Kenneth S. Pitzer Center for Theoretical Chemistry, University of California, Berkeley, California 94720} 

\date{\today}
\begin{abstract}
We present an integral transformation capable of extracting the moments of arbitrary Paley-Wiener entire functions $f(z)$ against a given spectral distribution $dP(\omega)$ from the short-time values of the correlation function $C(t)$. The transformation reads
\[
\int_\mathbb{R} f(\omega) dP(\omega) = \int_0^\infty e^{-s}\left[\frac{1}{2\pi i}\oint_\gamma f\left(\frac{s}{z}\right)\frac{C(-iz)}{z}dz\right]ds.
\]
It is proven to be valid for correlation functions that can be extended analytically to the entire complex plane with the possible exception of two branch cuts on the imaginary axis. If the analytic continuation exists only for a strip $|\mathrm{Im}(z)| < \tau_0$, then the integral transformation remains valid for all Paley-Wiener functions obtained by Fourier-Laplace transforming a compactly supported distribution, with the support included in the interval $(-2\tau_0, 2\tau_0)$.  Finally, if the support of the distribution is  contained in the interval $(-\tau_0, \tau_0)$, then the generalized moment can be evaluated from the short-time values of the correlation function exponentially fast.
\end{abstract}
\keywords{spectral functions, correlation functions, moment problem, analytic continuation}
\maketitle

\section{Introduction} 

Several authors have investigated the possibility of reconstructing spectral functions from imaginary-time correlation functions by inverting a real Laplace transform\cite{Gub91, Jar96, Gal94, Kim97, Kim98, Kri99, Kri01, Rab00, Rab02} or solving a Hamburger moment problem.\cite{Pre04, Pre05} It is assumed that the input information is represented either by the correlation function $C(t)$ for imaginary values of $t$ in a limited interval or, more strongly, by the values of $C(t)$ in some open disk about the origin in the complex plane. That the information in imaginary-time is greatly supplemented by its real-time counterpart, in as much as complementary aspects of the spectrum are described, has been recognized a while ago by Kim, Doll, and Freeman\cite{Kim98} as well as Krilov and Berne.\cite{Kri99}   My own effort to produce a path integral technique capable of furnishing information in an entire open disk\cite{Pre05} has been motivated not only by the physical insight obtained from the aforementioned research, but also by an intriguing observation of Lyness\cite{Lyn68} that some problems that are numerically unstable on the real line become numerically stable when formulated in the complex plane. Lyness' most explicit example was the problem of numerical differentiation, which he proved to be stable if performed by means of Cauchy's integral formula.   

Although the specific autocorrelation functions vary, they can always be casted in the form of a Fourier transform of a probability measure (see, for example, Ref.~\onlinecite{Pre05})
\begin{equation}
\label{eq:1a}
C(t) = \int_{\mathbb{R}} e^{i\omega t} dP(\omega). 
\end{equation}
In some formulations, the actual measure can be signed, but the negative and positive parts are always finite and, by linearity, no loss of generality is incurred by assuming $dP(\omega)$ to be a positive measure.  $C(t)$ is also called the characteristic function of $dP(\omega)$, whereas $C(-it)$, if it exists, constitutes the moment generating function. In many cases of practical importance, the imaginary-time correlation function $C(-it)$ is finite for $t$ in some range $(-\tau_0, \tau_0)$. $\tau_0$ is usually proportional to $\hbar \beta$ and is  rarely infinite. Because of the absolute integrability, it can easily be argued that $C(t)$ is in fact analytic in the entire strip with $|\mathrm{Im}(z)| < \tau_0$. Baym and Mermin\cite{Bay61} were perhaps the first to point out that this well-known fact in statistics automatically implies that the correlation function can be reconstructed from its imaginary-time values by analytic continuation. The latter quantities are normally assumed to be easier to obtain, for example by path integral Monte Carlo techniques. 

Analytic continuation is, of course, a numerically unstable procedure. The quality of the reconstructed spectral functions depends tremendously on the quality and the ``form'' in which the input information is given. In this paper, we assume that the information is available in the form of the values of the correlation function in an open disk $\{z\in \mathbb{C}:|z| < \tau_0\}$ about the origin, a disk  included in the region of analyticity of $C(t)$ (that is, only short-time information is available). By virtue of Lyness' observation, a Hamburger moment problem can be set up, with the moments of the spectral function computed by means of a contour integral. If $\gamma$ is a Jordan curve about the origin and situated in the region of analyticity of $C(t)$, we obtain 
\begin{equation}
\label{eq:2a}
\mu_k = \int_\mathbb{R} \omega^k dP(\omega) = \left. \frac{d^k}{dt^k}C(-it)\right|_{t = 0} = \frac{k!}{2\pi i} \oint_\gamma \frac{C(-iz)}{z^{k+1}}dz. 
\end{equation}
Tagliani\cite{Tag94} has employed this same observation to good effect  in order to substitute an inverse Hamburger moment problem for an inverse real Laplace transform. Given the known instability of the moment problem when monomials are utilized,\cite{Tyr94, Bek00} a natural question that arises is whether or not we can compute moments of more suitable functions, for example polynomials that are orthogonal with respect to some other measure or more general complex entire functions. It is the purpose of the present work to show that this is possible in several cases. 

\section{Computing generalized moments of spectral functions} 
 
I have been experimenting for some time with a modification of Eq.~(\ref{eq:2a}) of the form
\begin{equation}
\label{eq:3a}
\int_\mathbb{R} \omega^k dP(\omega) = \int_0^\infty e^{-s}\left[\frac{1}{2\pi i} \oint_\gamma \left(\frac{s}{z}\right)^k \frac{C(-iz)}{z}dz\right]ds. 
\end{equation}
Eq.~(\ref{eq:3a}) is equivalent to Eq.~(\ref{eq:2a}) because $k! = \int_0^\infty s^k e^{-s}ds$. Nevertheless, it allows for improved accuracy in evaluating moments of polynomials that are orthogonal with respect to another probability distribution, perhaps one that closely resembles $dP(\omega)$. Indeed, if $f(z) = \sum_{k = 0}^n a_k z^k$, then one can define the function
\begin{equation}
\label{eq:4a}
\mathcal{P}_f(s) = \frac{1}{2\pi i} \oint_\gamma f\left(\frac{s}{z}\right) \frac{C(-iz)}{z}dz
\end{equation}
and use linearity to justify the equality
\begin{equation}
\label{eq:5a}
\int_0^\infty e^{-s} \mathcal{P}_f(s) ds = \int_\mathbb{R} f(\omega) dP(\omega). 
\end{equation}
The catch here is that the polynomial $f(z)$ may actually be known in the factorized form, which is numerically stable. Then, the Gauss-Laguerre quadrature rule of appropriate order can be utilized to compute moments of arbitrary polynomials in a stable fashion. Notice that $\mathcal{P}_f(s)$ is a polynomial of degree $n$ as well and is exactly integrated by the Gauss-Laguerre quadrature rule if the number of knots is greater than $[n/2]+1$. Modified moments against orthogonal polynomials represent a significantly less redundant way of storing information. As pointed out by Sack and Donovan, \cite{Sac71} they may enable the computation of the quadrature knots and weights of the Gauss quadrature with weight function $dP(\omega)$. This quadrature rule defines a measure made up of point masses that can be shown to converge weakly to $dP(\omega)$ (see, e.g., Ref.~\onlinecite{Pre04} and Theorem~3.12 in Chapter~2 of Ref.~\onlinecite{Dur96}). 

Nevertheless, it is tempting  to utilize Eq.~(\ref{eq:5a}) together with more general analytic functions that are entire on $\mathbb{C}$. For example, a symmetric positive function $f(z)$ that decreases to the right and left on the real line, such as the Gaussian $f(z) = \exp(-z^2/2)/\sqrt{2\pi}$, constitutes the perfect example for constructing delta sequences that might provide the value of the spectrum at a point of continuity. Starting with such a function, we can obtain a complete description of the spectrum in the form 
\begin{equation}
\label{eq:6a}
\frac{1}{\sigma}\int_\mathbb{R} f\left(\frac{\omega - \omega_0}{\sigma}\right)dP(\omega),
\end{equation} 
for any level of resolution $\sigma > 0$ desired. Surely, we would like to utilize entire functions $f(z)$ with tails that decay fast on the real line. In the following section, we will show that this is always possible if the measure $dP(\omega)$ is compactly supported. 

Unfortunately, for the more interesting cases, the measure is not compactly supported. Its tail only decays as an exponential of the form $\exp(-\tau_0|\omega|)$, for some $\tau_0 > 0$. A suggestive example is given by the Miller, Schwartz, and Tromp flux-flux spectral function for the free particle,\cite{Mil83} which has the continuous distribution
\begin{equation}
\label{eq:7a}
\bar{C}_F(\omega) =\frac{1}{\beta h} \frac{|\omega| \hbar \beta}{2\pi} K_1\left(\frac{|\omega| \hbar \beta}{2}\right),
\end{equation}
where $K_1(x)$ denotes the respective modified Bessel function of the second kind. Setting $\tau_0 = \hbar\beta/2$, the tail of the distribution behaves roughly as $|\omega|^{-1/2}\exp(-\tau_0 |\omega|)$.  The consequence is that the distribution does not integrate  exponentials of order greater than or equal to $\tau_0$ and, therefore, the correlation function is not defined for purely imaginary times of the form $\{i\tau: |\tau| \geq \tau_0\}$. The actual correlation function reads
\begin{equation}
\label{eq:8a}
C_F(t) = \frac{1}{\beta h} \frac{(\beta \hbar / 2 )^2}{\left[t^2 + (\beta \hbar / 2 )^2\right]^{3/2}},  
\end{equation}
and is seen to exhibit two branch cuts on the imaginary axis, branch cuts that form the set $\{z\in \mathbb{C} : z = i\tau, |\tau| \geq \tau_0\}$.

Nevertheless, it is the case with this correlation function and, perhaps, with many other correlation functions of practical importance that they are analytic everywhere  with the possible exception of two branch cuts. Under this assumption, in Section~IV, we prove that $\mathcal{P}_f(s)$ is integrable against $e^{-s}ds$ and that Eq.~(\ref{eq:5a}) is true for all entire functions of Paley-Wiener type. These functions are given by the Fourier-Laplace transforms of some compactly supported distribution $\hat{f}(\kappa)$, i.e.,
\begin{equation} 
\label{eq:9a}
f(z) = \int_\mathbb{R}\hat{f}(\kappa)e^{i\kappa z}d\kappa. 
\end{equation}
In the most general case, the correlation function can be extended only to a strip of analyticity defined by the inequality $|\mathrm{Im}(z)| < \tau_0$. In such cases,  Eq.~(\ref{eq:5a}) is valid for those Paley-Wiener entire functions which are the Fourier-Laplace transform of distributions with the support included in the compact interval $(-2\tau_0, 2\tau_0)$. 

Having established the physical and mathematical motivation, we dedicate the remainder of the paper to the actual proofs.

\section{The compactly supported case}

\begin{1}
\label{Th:1}
Let $dP(\omega)$ be a probability measure on $\mathbb{R}$ such that its characteristic function is analytic at the origin and let $\tau_0 > 0$ be the radius of convergence of the Taylor series about the origin ($\tau_0$ may be infinite).   Let $f : \mathbb{C} \to \mathbb{C}$ be an entire analytic function, the real restriction of which increases slower than any exponential, i.e., for any $\epsilon > 0$ there exists a constant $M > 0$  such that $ | f(x)| < M e^{\epsilon |x|}$, $ \forall \; x \in \mathbb{R}$. Then $\mathcal{P}_f(s)$ is an entire function. If $dP(\omega)$ is compactly supported, then 
\begin{equation}
\label{eq:1c}
\int_0^\infty e^{-s} \mathcal{P}_f(s)ds = \int_{\mathbb{R}} f(\omega)dP(\omega). 
\end{equation}
\end{1}

\emph{Proof.}  By hypothesis, the Taylor series of $f$ about the origin $f(z) = \sum_{k \geq 0} a_k z^k$ is absolutely convergent on the whole complex plane, whereas the Taylor series of the moment generating function $C(-it) = \sum_{k \geq 0} \mu_k t^k / k!$ is convergent in an open disk of radius $\tau_0$. It follows that the contour integral expressed by Eq.~(\ref{eq:4a}) is convergent for all $s \in \mathbb{C}$. By Cauchy's theorem and the uniform convergence of the partial sums on the contour $\gamma$, the result is the convolution series
\begin{equation}
\label{eq:2c}
\mathcal{P}_f(s) = \sum_{k = 0}^\infty a_k \mu_k \frac{s^k}{k!}. 
\end{equation}  
$\mathcal{P}_f(s)$ is an entire function since it is a power series convergent for all $s \in \mathbb{C}$.

We treat the remainder of the theorem by starting with some preparatory material. The measure $dP(\omega)$ integrates to a finite number any function of exponential order $\tau < \tau_0$. Indeed, 
\begin{eqnarray}
\label{eq:3c} \nonumber
\int_\mathbb{R} |f(\omega)| dP(\omega) \leq M \int_\mathbb{R} e^{\tau |\omega|} dP(\omega) \leq M \int_\mathbb{R} (e^{\tau \omega}+e^{-\tau \omega}) dP(\omega)\\ = M [C(i\tau) + C(-i\tau)] < \infty. 
\end{eqnarray}
Thus, the right-hand side of Eq.~(\ref{eq:1c}) is finite. 

Let $g : \mathbb{C} \to \mathbb{C}$ be defined by
\begin{equation}
\label{eq:4c}
g(z) = \sum_{k = 0}^\infty a_k \frac{z^k}{k!}. 
\end{equation}
$g(z)$ is also entire and, for any $\epsilon > 0$, there is $M_\epsilon > 0$ such that $|g(z)| < M_\epsilon \exp(\epsilon |z|)$. These assertions are established at once by the sequence of inequalities
\begin{eqnarray}
\label{eq:5c} \nonumber
|g(z)| \leq \sum_{k \geq 0} |a_k| \frac{|z|^k}{k!} = \sum_{k \geq 0} |a_k|\epsilon^{-k} \frac{|z\epsilon|^k}{k!}  \leq 
\left(\sum_{k \geq 0} |a_k|\epsilon^{-k}\right) \\ \times \left(\sum_{k \geq 0}\frac{|z\epsilon|^k}{k!}\right)  = \left(\sum_{k \geq 0} |a_k| \epsilon^{-k} \right)e^{\epsilon |z|}
\end{eqnarray}
and the finitude of the last series in parenthesis (the series, which is a power series, is finite by the absolute convergence of the Taylor series of $f$ about the origin for an arbitrary radius). In fact, Eq.~(\ref{eq:5c}) also establishes that the absolute series $\sum_{k \geq 0} |a_k| {|z|^k}/{k!}$ is of exponential order $\epsilon$, for all $\epsilon > 0$. Thus, the series $g(\omega s) = \sum_{k \geq 0} a_k (\omega s)^k / k!$ can be integrated term by term against both measures $e^{-s}ds$ and $dP(\omega)$. We obtain
\begin{equation}
\label{eq:6c}
\mathcal{P}_f(s) = \sum_{k = 0}^\infty a_k \frac{s^k}{k!} \left[\int_\mathbb{R} \omega^k dP(\omega)\right] = \int_\mathbb{R} \left[\sum_{k = 0}^\infty a_k \frac{s^k}{k!}  \omega^k\right] dP(\omega) = \int_\mathbb{R} g(\omega s) dP(\omega)
\end{equation}
and 
\begin{eqnarray}
\nonumber
\label{eq:7c} &&
f(\omega) = \sum_{k = 0}^\infty a_k \omega^k = \sum_{k = 0}^\infty a_k \omega^k \frac{1}{k!}\left[\int_0^\infty s^k e^{-s}ds \right] \\ && = \int_0^\infty e^{-s}\left[\sum_{k = 0}^\infty a_k  \frac{(\omega s)^k}{k!} \right] ds = \int_0^\infty e^{-s} g(\omega s) ds.
\end{eqnarray}

The reader may already see why Eq.~(\ref{eq:1c}) could be true. It boils down to proving a change in the order of integration. Indeed, by Eq.~(\ref{eq:7c}), 
\begin{equation}
\label{eq:8c}
\int_\mathbb{R}f(\omega)dP(\omega) = \int_\mathbb{R} \left[\int_0^\infty e^{-s} g(\omega s) ds\right]dP(\omega), 
\end{equation}
whereas by Eq.~(\ref{eq:6c}) and assuming $\mathcal{P}_f$ is integrable against $e^{-s}ds$, 
\begin{equation}
\label{eq:9c}
\int_0^\infty e^{-s}\mathcal{P}_f(s)ds = \int_0^\infty e^{-s}\left[\int_\mathbb{R} g(\omega s) dP(\omega)\right]ds.
\end{equation}
The change in the order of integration as well as the integrability of $\mathcal{P}_f(s)$ could be justified by the Fubini-Tonelli theorem provided that the kernel $g(\omega s)$ were integrable against the product measure. This is true for  measures $dP(\omega)$  supported in some arbitrary interval $[-r, r]$, with $r > 0$. Since in this case $|\mu_k| < 2r^k$, we calculate
\begin{equation}
\label{eq:10c}
\int_0^\infty e^{-s}ds \int_{\mathbb{R}} dP(\omega) |g(\omega s)| \leq \sum_{k = 0}^\infty |a_k||\mu_k| \leq 2 \sum_{ k = 0}^\infty |a_k| r^k < \infty, 
\end{equation}
again by the absolute convergence of the Taylor series of $f(z)$. The proof of the theorem is completed. \hspace{\stretch{1}}$\Box$

A quick look at Eq.~(\ref{eq:10c}) shows that the ease with which we have demonstrated the validity of Eq.~(\ref{eq:1c}) in such a general setting is due to the convergence of the series
\begin{eqnarray}
\label{eq:11c} \nonumber
F(\theta) = \sum_{k = 0}^\infty a_k \mu_k \theta^k = \sum_{k = 0}^\infty a_k \theta^k \int_{\mathbb{R}}  \omega^k dP(\omega) = \int_{\mathbb{R}} \sum_{k = 0}^\infty a_k   (\theta \omega)^k dP(\omega) \\  = \int_{\mathbb{R}} f(\theta \omega) dP(\omega), \quad \forall \; \theta \in \mathbb{C}. 
\end{eqnarray}
That the Taylor series of $f(\theta z)$ can be integrated term by term follows from the fact that  the domain of integration is some finite interval $[-r,r]$. Thus, the right-hand side of Eq.~(\ref{eq:11c}) is always finite because $f(\theta \omega)$ is bounded on $[-r, r]$, since it is continuous. Then, the uniform convergence of the Taylor series justifies the change of order between integration and summation and establishes the finitude of the left-hand side series. 

If the measure $dP(\omega)$ is not compactly supported, we cannot expect the left-hand side series to be convergent for all $\theta \in \mathbb{C}$, unless the function $f(z)$ is rather trivial (e.g., a polynomial). In fact, the series may not be convergent at all for the more interesting functions, such as $f(z) = \exp(-z^2) = \sum_{k \geq 0} (-z^2)^k / k!$. Indeed, if $dP(\omega) = \exp(-|\omega|)/2d\omega$, the even moments of which are $\mu_{2k} = (2k)!$, we compute
\[
F(\theta) = \sum_{k = 0}^\infty a_k \mu_k \theta^k = \sum_{k = 0}^\infty (-1)^k \frac{(2k)!}{k!}\theta^{2k},
\]
which is divergent  for all $\theta \neq 0$ (by the ratio test). On the other hand, if $f(z) = \exp(iz) = \sum_{k \geq 0} (iz)^k/k!$, an interesting thing happens. The series 
\begin{equation}
\label{eq:12c}
F(\theta) = \sum_{k = 0}^\infty a_k \mu_k \theta^k = \sum_{k = 0}^\infty  (i\theta)^{2k} = \frac{1}{1+\theta^2}, \quad  |\theta| < 1,
\end{equation}
is convergent if $|\theta| < 1$, but divergent if $|\theta| > 1$. Nevertheless, the integral
\begin{equation}
\label{eq:13c}
F(\theta) =  \int_{\mathbb{R}} f(\theta \omega) dP(\omega) = \frac{1}{2} \int_{\mathbb{R}}  e^{-|\omega|} e^{i\theta \omega} d\omega = \frac{1}{1+\theta^2}, \quad  |\mathrm{Im}(\theta)| < 1,
\end{equation}
is convergent for all $\theta$ of imaginary part in the interval $(-1,1)$.

Clearly, in this case, $F(\theta)$ is just the correlation function of the measure $dP(\omega)$. In the introduction, we have already mentioned that $F(\theta)$ is analytic in the strip defined by the equation $|\mathrm{Im}(\theta)| < \tau_0$ (for the example described by Eq.~\ref{eq:13c}, $\tau_0 = 1$). Given the relevance of the convergence of the series expansion of $F(\theta)$ in the proof of Theorem~\ref{Th:1}, it makes sense to restrict our attention to entire functions that exhibit such properties. With respect to their ability of recovering information via delta sequences, our intuition says that functions like $f(z) = \exp(-z^2)$ are, perhaps, of little help. Functions like $f(z) = \sin(z)/z$, which exhibit properties similar to $f(z) = \exp(iz)$, may actually do the job. In the next section, we show that the latter assertion is true, indeed, under the assumption that the correlation function of $dP(\omega)$ can be analytically continued to the whole complex plane save two branch cuts on the imaginary axis. 

\section{The general case}
For the remainder of the present paper, we shall assume that the function $f(z)$ and the measure $dP(\omega)$ satisfy the conditions of Theorem~\ref{Th:1} save the compactness of the support of the measure. In fact, the functions $f(z)$ will be restricted to a smaller subset, in agreement with the opinion expressed in the last paragraph of the preceding section. The class is provided by the following definition. 
\begin{2d}
\label{Def:2}
An entire function $f : \mathbb{C} \to \mathbb{C}$ is said to be of Paley-Wiener type (or simply Paley-Wiener function) if it is the Fourier-Laplace transform of some compactly supported distribution $\hat{f}(\kappa)$, i.e.,
\begin{equation}
\label{eq:1df}
f(z) = \int_\mathbb{R} \hat{f}(\kappa) e^{i\kappa z}d\kappa.
\end{equation} 
\end{2d}
The distributional version of the Paley-Wiener theorem asserts that if the distribution is supported in the interval $[-B, B]$, with $B > 0$, there exist positive constants $C$ and $N$ such that
\begin{equation}
\label{eq:1ds}
|f(z)| \leq C (1 + |z|)^N e^{B |\mathrm{Im}(z)|}.
\end{equation}
(The converse is also true). Thus, the function $f(z)$ has at most a polynomial growth for real arguments. In fact,  a Paley-Wiener function $f(z)$ has at most a polynomial growth in any strip $|\mathrm{Im}(z)|< \tau$,  with $\tau > 0$. 

\begin{1p}
\label{Pr:1}
Let $f: \mathbb{C} \to \mathbb{C}$ be a Paley-Wiener function and let $S$ be the strip $S = \{\theta \in \mathbb{C}:\left|\mathrm{Im}(\theta)\right| < \tau_0 / B\}$. Then, 
\begin{equation}
\label{eq:1d}
\int_{\mathbb{R}} \left|f(\theta \omega)\right| dP(\omega) < \infty, \quad \forall \; \theta \in S.
\end{equation} 
The function $F(\theta) : S \to \mathbb{C}$ defined by
\begin{equation}
\label{eq:2d}
F(\theta) = \int_{\mathbb{R}} f(\theta \omega) dP(\omega)
\end{equation}
is analytic. More specifically, the Taylor series of $f(\theta \omega)$ (regarded as a function of $\theta$) about any real number $\theta_r$ can be integrated term by term and produces the Taylor series of $F(\theta)$ about $\theta_r$, a series having a radius of convergence of at least $\tau_0/B$. 
\end{1p}
\emph{Proof.} We have already mentioned during the proof of Theorem~\ref{Th:1} that the measure $dP(\omega)$ integrates all functions of exponential order less than $\tau_0$ (according to Eq.~\ref{eq:3c}). Thus, according to Eq.~(\ref{eq:1ds}),  the integral in Eq.~(\ref{eq:1d}) is absolutely convergent for all $\theta$ with $B|\mathrm{Im}(\theta)| < \tau_0$.  Even more, if $K$ is a compact set included in the strip $S$, we have $B |\mathrm{Im}(\theta)| < \tau_1 < \tau_0$ and $|\theta| < \tau_2$ for some $\tau_1, \tau_2 > 0$ and for all $\theta \in K$ [by continuity, $\mathrm{Im}(\theta)$ and $|\theta|$ attain their extrema at some points in $K$, a compact set which is properly included in $S$]. Then the inequality
\begin{eqnarray}
\label{eq:3df}\nonumber
\int_{\mathbb{R}}|f(\theta \omega)|dP(\omega) \leq C \int_{\mathbb{R}}(1 + |\theta \omega|)^N e^{B |\mathrm{Im}(\theta)|\omega}dP(\omega) \\ \leq C \int_{\mathbb{R}}(1 + \tau_2|\omega|)^N e^{\tau_1\omega}dP(\omega) < \infty
\end{eqnarray}
tells us that  the integral defined by Eq.~(\ref{eq:1d}) is not only finite, but also uniformly bounded on each compact set. By the continuity of the integrand and the dominated convergence theorem, it follows that $F(\theta)$ is continuous on $S$. 

Let $\theta_r \in \mathbb{R}$ be arbitrary and let $f^{(k)}(z)$ denote the $k$-th derivative of $f(z)$. Since $f(z)$ is an entire function, the Taylor series of $f(\theta \omega)$ regarded as a function of $\theta$
\begin{equation}
\label{eq:3d}
f(\theta \omega) = \sum_{k = 0}^\infty f^{(k)}(\theta_r \omega) \frac{\omega^k}{k!} (\theta - \theta_r)^k  
\end{equation} 
is convergent on the entire complex plane, for all $\omega \in \mathbb{R}$. By Cauchy's integral formula (let $\gamma$ be a Jordan curve encircling the point $\theta_r$),
\[
f^{(k)}(\theta_r \omega) \omega^k = \left. \frac{d^k f(\theta \omega)}{d\theta^k}\right|_{\theta = \theta_r} = \frac{k!}{2\pi i} \oint_\gamma \frac{f(z\omega)}{(z - \theta_r)^{k+1}}dz,
\]
so that the Taylor series now reads
\begin{equation}
\label{eq:4d}
f(\theta \omega) = \sum_{k = 0}^\infty \frac{(\theta - \theta_r)^k }{2\pi i} \oint_\gamma \frac{f(z\omega)}{(z - \theta_r)^{k+1}}dz. 
\end{equation}

Given an arbitrary $\theta$ such that $|\theta - \theta_r| < \tau_0/B $ (with $\theta_r\in \mathbb{R}$, as in the hypothesis of the problem), we  choose the contour $\gamma$ to be a circle centered about $\theta_r$ and of radius $\tau > 0$, with $\tau$ such that $|\theta - \theta_r|< \tau < \tau_0/B $. Then the integrability term by term against the measure $dP(\omega)$ of the series given by Eq.~(\ref{eq:4d}) follows from the Fubini theorem once we prove the absolute integrability against the product measure (of $dP(\omega)$, the counting measure, and the Lebesgue measure for the polar parameterization of $\gamma$). Using Tonelli's theorem to justify the arbitrary order of integration, we compute 
\begin{eqnarray*} &&
\int_{\mathbb{R}}dP(\omega)\sum_{k = 0}^\infty \int_0^{2\pi}d\phi \frac{|\theta - \theta_r|^k}{2\pi}  \frac{\left| f\left[(\theta_r+\tau e^{i\phi})\omega \right]\right|}{|\theta_r + \tau e^{i\phi} - \theta_r|^{k+1}} 
\\ && = 
 \frac{1}{2\pi \tau }\left(\sum_{k = 0}^\infty   \frac{|\theta - \theta_r|^k}{\tau ^{k}}\right) \int_0^{2\pi}d\phi \left\{\int_{\mathbb{R}}dP(\omega)\left| f\left[(\theta_r+\tau e^{i\phi})\omega \right]\right|\right\}
 \\ && = 
\frac{1}{2\pi\left(\tau - |\theta - \theta_r|\right)} \int_0^{2\pi}d\phi \left\{\int_{\mathbb{R}}dP(\omega)\left| f\left[(\theta_r+\tau e^{i\phi})\omega \right]\right|\right\}. 
\end{eqnarray*}
The geometric series is convergent because $|\theta - \theta_r| < \tau$. On the other hand, since $\left|\mathrm{Im}(\theta_r+\tau e^{i\phi})\right| \leq \tau < \tau_0/B$, the integral against $dP(\omega)$ is convergent for all $\phi \in [0, 2\pi)$. In fact, given that the circle constituting the contour is a compact set included in $S$, the integral against $dP(\omega)$, as a function of $\phi$, is uniformly bounded. It follows that the integral against $d\phi$ is finite and the term-by-term integration of the Taylor series is established. We therefore have
\begin{equation}
\label{eq:5d}
F(\theta) = \sum_{k = 0}^\infty  \left[ \int_{\mathbb{R}} f^{(k)}(\theta_r \omega)\omega^k dP(\omega)\right]\frac{(\theta - \theta_r)^k}{k!}.
\end{equation} 
 Since $\theta_r \in \mathbb{R}$ and $\theta$ with $|\theta - \theta_r| < \tau_0/B $ are otherwise arbitrary, the proof is concluded. \hspace{\stretch{1}}$\Box$ 
 
The preceding proposition tells us that the Paley-Wiener functions borrow many of the properties of $f(z)= \exp(iz)$. The latter function generates the correlation function itself
\[
C(\theta) = \int_\mathbb{R} e^{i\theta \omega} dP(\omega).
\]
Thus, we obtain once again that the correlation function  is  analytic on the entire strip $|\mathrm{Im}(\theta)| < \tau_0$ (in this case $B = 1$). The function $F(\theta)$ can be expressed in terms of the correlation function as follows:
\begin{equation}
\label{eq:6df}
F(\theta) =  \int_\mathbb{R} \left[\int_\mathbb{R}\hat{f}(\kappa)e^{i\kappa\theta\omega}d\kappa\right] dP(\omega) = \int_\mathbb{R} \hat{f}(\kappa)\left[\int_\mathbb{R}e^{i\kappa\theta\omega}dP(\omega)\right]d\kappa = \int_\mathbb{R} \hat{f}(\kappa)C(\theta \kappa)d\kappa.
\end{equation}
If $|\theta| < \tau_0/B$, the formal change in  the order of integration can be rationalized by means of Eq.~(\ref{eq:5d}), as applied for $\exp(iz)$ and $f(z)$ about the origin. Indeed, the series 
\[
C(\theta \kappa) = \sum_{j = 0}^\infty \kappa^j \left[ \int_{\mathbb{R}}  i^j\omega^j dP(\omega)\right]\frac{\theta^j}{j!}
\]
converges absolutely for all $|\kappa \theta| < \tau_0$. As a function of $\kappa$, the series converges in the $C^\infty$ topology on the  open interval $|\kappa| < \tau_0/|\theta|$ (the case with $\theta = 0$ is trivially true). This open interval contains the support of the distribution $\hat{f}(\kappa)$, because $|\theta| < \tau_0/B$. By continuity of the distribution as a linear functional, 
\begin{eqnarray*}
\int_\mathbb{R} \hat{f}(\kappa)C(\theta \kappa)d\kappa = \sum_{j = 0}^\infty \left[\int_\mathbb{R} \hat{f}(\kappa) \kappa^j d\kappa\right] \left[ \int_{\mathbb{R}} i^j\omega^j dP(\omega)\right]\frac{\theta^j}{j!} \\ =\sum_{j = 0}^\infty  \left[ \int_{\mathbb{R}} f^{(j)}(0)\frac{(\omega \theta)^j}{j!} dP(\omega)\right]. 
\end{eqnarray*}
The order of the summation and integration can be changed again by means of Eq.~(\ref{eq:5d}), as applied to  $f(z)$ about the origin. Then the interior series is seen to sum to $f(\theta \omega)$ and its integral equals $F(\theta)$, indeed.  

We can extend the validity of Eq.~(\ref{eq:6df}) from the disk $| \theta| < \tau_0 /B$ to the strip $S$ by analytic continuation. Indeed, the left-hand side has been proven to be analytic in the strip $S$. It will coincide with the right-hand side expression on $S$ provided that the latter is analytic in $\theta$ on $S$. The right-hand side expression  is well-defined because $C(\theta \kappa)$, as a function of $\kappa$, is analytic in the open set (regarded as a subset in $\mathbb{R}$) defined  by the equation  $|\mathrm{Im}(\theta \kappa)| = |\kappa| |\mathrm{Im}(\theta)| < \tau_0$. If $\theta \in S$, then this open set contains the support of $\hat{f}(\kappa)$  because $|\mathrm{Im}(\theta)| < \tau_0/B$. By a similar argument, the first order derivative of the right-hand side expression, namely $\int_\mathbb{R} \hat{f}(\kappa)\kappa C^{(1)}(\theta \kappa)d\kappa$,  also exists for all $\theta \in S$.

To continue, let us notice that it is enough to show that, for each $\theta \in S$, there exists a compact set $K \in S$ that contains all the complex points of the form $z\kappa$, with $z$ in some neighborhood $U$ of $\theta$ and $\kappa$ in some precompact open set containing $[-B, B]$. In this case, the second derivative of $C(z \kappa)$ against $z$, i.e., the function $C^{(2)}(z\kappa)\kappa^2$, is bounded as absolute value by some constant $M > 0$ (by continuity). If $\theta_n \to \theta$, the Taylor series with remainder says that, for $n$ large enough,
\[
\left|\frac{C(\theta \kappa) - C(\theta_n \kappa)}{\theta - \theta_n} - C^{(1)}(\theta \kappa)\kappa \right|  \leq \frac{1}{2} \sup_{z \in \overline{U}}\left|C^{(2)}(z\kappa)\kappa^2\right||\theta - \theta_n| \leq \frac{M}{2} |\theta - \theta_n|. 
\]
Thus, the convergence $[C(\theta \kappa) - C(\theta_n \kappa)]/(\theta - \theta_n) \to C^{(1)}(\theta \kappa)\kappa$ is uniform with respect to $\kappa$ in an open set containing the support of $\hat{f}(\kappa)$. However, a similar argument with $C(\theta \kappa)$ replaced by $C^{(j)}(\theta \kappa)\theta^j$ shows that, for every $j \geq 1$,
\[\frac{d^j}{d \kappa^j}\frac{C(\theta \kappa) - C(\theta_n \kappa)}{\theta - \theta_n} \to \frac{d^j}{d \kappa^j} \left[C^{(1)}(\theta \kappa)\kappa\right]\]
uniformly on the same precompact open set containing the support of $\hat{f}(\kappa)$. It follows that the convergence $[C(\theta \kappa) - C(\theta_n \kappa)]/(\theta - \theta_n) \to C^{(1)}(\theta \kappa)\kappa$ happens in the $C^{\infty}$ topology on that set and therefore
\begin{eqnarray*}
\lim_{n \to \infty} \int_{\mathbb{R}} \hat{f}(\kappa) \frac{C(\theta \kappa) - C(\theta_n \kappa)}{\theta - \theta_n} d\kappa = \int_{\mathbb{R}}  \hat{f}(\kappa) \lim_{n \to \infty} \left[ \frac{C(\theta \kappa) - C(\theta_n \kappa)}{\theta - \theta_n}\right] d\kappa \\ = \int_{\mathbb{R}}  \hat{f}(\kappa) C^{(1)}(\theta \kappa)\kappa d\kappa.
\end{eqnarray*}
Since the last quantity exists, we obtain that $\int_\mathbb{R} \hat{f}(\kappa)C(\theta \kappa)d\kappa$ is differentiable at any point $\theta \in S$. It therefore defines an analytic function on $S$, which must coincide with $F(\theta)$.

Finding such a compact set $K$ for a given $\theta \in S$ is not difficult. Let $\epsilon = [\tau_0/B - |\mathrm{Im}(\theta)|]/2$. Of course, $\epsilon > 0$, since $\theta \in S$. Because $0<B(\tau_0/B - \epsilon) < B(\tau_0/B - \epsilon/2)$, there is $\eta > 0$ such that $(B+\eta)(\tau_0/B - \epsilon) < B(\tau_0/B - \epsilon/2)$. Consider the set $K$ made up of the numbers of the form $\kappa z$ with $\kappa \in (-B-\eta, B+\eta)$ and $|z - \theta| < \epsilon$. Any element $\kappa z$ of $K$ satisfies the inequalities $|\mathrm{Im}(\kappa z)| = |\kappa| |\mathrm{Im}(z)| < (B+\eta)(\tau_0/B-\epsilon) < \tau_0 - B\epsilon/2 $ as well as $|\mathrm{Re}(\kappa z)| = |\kappa| |\mathrm{Re}(z)| < (B+\eta)[|\mathrm{Re}(\theta)| + \epsilon]$. If we replace $K$ with its closure, the bounds transform from strict inequalities to inequalities, but the set is still properly contained in $S$. By the discussion in the preceding paragraph, Eq.~(\ref{eq:6df}) is valid on $S$. 

Nevertheless, the same technique can be applied to analytically continue $F(\theta)$ on a larger connected domain that includes the open disk $|\theta|< \tau_0/B$. Indeed, there was nothing special about $S$ except for the existence of an appropriate compact set $K$ associated with a point $\theta$. Assume now that $C(\theta)$ can be extended analytically on the whole complex plane except perhaps for the subset $\{z \in \mathbb{C}: z = iy, |y| \geq \tau_0\}$. Then for any $\theta \in \mathbb{C}$ with $\eta = \mathrm{Re}(\theta) \neq 0$, the  set $K = \{ z \kappa \in \mathbb{C} : \kappa \in (-B-1, B+1), |z - \theta| < |\eta / 2|\}$ is included in the domain of analyticity of $C(\theta)$. Indeed, one notices that the inequality $|\mathrm{Re}(z\kappa)| = |\kappa| \mathrm{Re}(z) > |\kappa|\eta/2$ prohibits the set $K$ from intersecting the forbidden region. Since $K$ is trivially bounded, its closure is compact. By the continuity of the function $\mathrm{Re}(z)$, the preceding inequality remains valid on the closure of $K$, perhaps with the sign $>$ replaced by $\geq$.  It follows  that $F(\theta)$, as now re-defined by Eq.~(\ref{eq:6df}), is analytic on the open set $\mathrm{Re}(\theta) \neq 0$.  We summarize the findings of the last two paragraphs in the following proposition.
\begin{3p} 
\label{Pr:2}
On $S = \{z\in \mathbb{C}: |\mathrm{Im}(z)| < \tau_0/B\}$, the following equality of analytic functions holds 
\begin{equation}
\label{eq:6ds}
F(\theta) = \int_\mathbb{R} \hat{f}(\kappa)C(\theta \kappa) d\kappa. 
\end{equation}
In addition, if $C(\theta)$ can be analytically continued to the whole complex plane except for the set $\{z \in \mathbb{C}: z = iy, |y| \geq \tau_0\}$, then $F(\theta)$ can be analytically continued  to the whole complex plane except for the set $\{z \in \mathbb{C}: z = iy, |y| \geq \tau_0/B\}$, by means of Eq.~(\ref{eq:6ds}). 
\end{3p}

The relation between $\mathcal{P}_f(s)$ and $F(\theta)$ can easily be established from the Taylor series expansion of $\mathcal{P}_f(s)$, which is given by Eq.~(\ref{eq:2c}). If $\gamma$ is any Jordan curve encircling the origin and enclosed in the disk about origin of radius $\tau_0 /B$, then the Cauchy integral formula and the uniform convergence of the Taylor series show that
\begin{equation}
\label{eq:6d}
\mathcal{P}_f(s) = \sum_{k = 0}^\infty a_k \mu_k \frac{s^k}{k!} = \frac{1}{2\pi i} \oint_\gamma \left(\sum_{k = 0}^\infty a_k \mu_k z^k \right) \left(\sum_{k = 0}^\infty \frac{1}{k!}\frac{s^k}{z^k} \right)\frac{dz}{z} = \frac{1}{2\pi i} \oint_\gamma z^{-1}F(z)e^{s/z}dz.
\end{equation}
However, after the summation of the series is done, the contour can be extended to the strip $S$ or to the whole domain of analyticity of $F$.

We now have all the tools to demonstrate the first theorem of the present section. 
\begin{2}
\label{Th:2}
Assume  $f$ is a Paley-Wiener function and $dP(\omega)$ is a probability measure, the correlation function of which can be analytically continued to the whole complex plane except for the set $\{z \in \mathbb{C}: z = iy, |y| \geq \tau_0\}$. Then the restriction of $\mathcal{P}_f(s)$ to $[0,\infty)$ is integrable against $e^{-s}ds$ and
\begin{equation}
\label{eq:8d}
\int_0^\infty e^{-s}\mathcal{P}_f(s) ds = \int_{\mathbb{R}} f(\omega) dP(\omega).
\end{equation}
\end{2}

\emph{Proof.} From the definition of $F(\theta)$, it follows the Eq.~(\ref{eq:8d}) is equivalent to
\[
 \int_0^\infty e^{-s}\mathcal{P}_f(s) ds= F(1)
\]
or, by Eq.~(\ref{eq:6d}), to 
\begin{equation}
\label{eq:9d}
\int_0^\infty e^{-s}\left[\frac{1}{2\pi i} \oint_\gamma z^{-1}F(z)e^{s/z}dz\right] ds = F(1).
\end{equation}
We can take the contour $\gamma$ to be the circle centered about $1/2$ and of radius $r = [1 + (\tau_0/B)^2]^{1/2}/2$. Indeed, the circle contains the origin in its interior because its radius is strictly greater than $1/2$. It is included in the domain of analyticity of $F(\theta)$ because the radius is strictly smaller than $[1/4 + (\tau_0/B)^2]^{1/2}$, the minimal distance to the forbidden set. 

If $z = x + 1/2 + iy \in \gamma$, we have $x^2 + y^2 = r^2$ and $(x+1/2)^2 + y^2 = r^2 + x + 1/4 \leq r^2 + r + 1/4 = (r+1/2)^2$. As such,
\[
\mathrm{Re}\left(1 - \frac{1}{z}\right) = 1- \frac{x+1/2}{(x+1/2)^2 + y^2} = \frac{r^2 -1/4}{r^2+x+1/4} \geq \frac{(\tau_0/B)^2}{(2r + 1)^2}  > 0.
\]
If $M > 0$ is an upper abound for $|z^{-1} F(z)|$ on the circle $\gamma$, then
\[
\left|e^{-s(1-1/z)}z^{-1}F(z)\right| \leq M \exp\{-s \mathrm{Re}(1-z^{-1})\} \leq M \exp\left\{-s \frac{(\tau_0/B)^2}{(2r + 1)^2}\right\}
\]
is uniformly bounded by a function that is integrable against the product measure. By Fubini-Tonelli's theorem, it follows that $F(s)$ is integrable against $e^{-s}ds$. By inverting the order of integration, we compute
\[
\frac{1}{2\pi i} \oint_\gamma z^{-1}F(z)\left[\int_0^\infty e^{-sz/(z-1)}ds\right] dz = \frac{1}{2\pi i} \oint_\gamma \frac{F(z)}{z-1} dz = F(1).
\]
The proof of the theorem is concluded. \hspace{\stretch{1}}$\Box$

An important corollary of  Theorem~\ref{Th:2} is the following statement asserting that the correlation function can be reconstructed pointwise. In this case, the functions $f(z)$ have the form $\exp(iz t)$ and are obtained by Fourier transforming point masses (Dirac functions) about $\kappa = t$. Setting $\mathcal{P}_f (s) = \mathcal{P}[f](s)$ for clarity and using $\star$ to denote the dummy variable $z$, we have
\begin{1c}
In the conditions of Theorem~\ref{Th:2}, 
\begin{equation}
\label{eq:10d}
C(t) = \int_0^\infty e^{-s}\mathcal{P}\left[e^{i\star t}\right](s) ds,
\end{equation}
for all $t \in \mathbb{R}$.
\end{1c}

Looking back at the proof of Theorem~\ref{Th:2}, we can see that the reason we require the existence of an analytic continuation of $F(\theta)$ is  that the quantity $\mathrm{Re}\left(1 - z^{-1}\right)$ becomes negative in a circle about $1/2$ of radius $1/2$. Thus, the Fubini-Tonelli theorem cannot be utilized if the contour $\gamma$ passes through the disk delimited by this circle. However, if the circle of radius $1/2$ centered about $1/2$ is properly contained in $S$, and this happens if  $1/2 < \tau_0/B$, we can always pick a circle centered about $1/2$ of radius strictly greater than $1/2$ that is still properly included in $S$. This circle contains the origin in its interior. On it, the function $\left|e^{-s(1-1/z)}\right|$ is bounded by $\exp(-\epsilon s)$ for some $\epsilon > 0$, whereas $|z^{-1}F(z)|$ is bounded by some $M > 0$.  Because the circle is contained in $S$, the analytic continuation of $F(\theta)$ outside $S$  is no longer needed. However, because  $B$ cannot be greater than $2\tau_0$, we lose the ability of constructing delta sequences by rescaling the argument of the function $f(z)$ (the information is band-limited). We obtain the following result.
\begin{3}
\label{th:3}
Regardless of whether an analytic continuation of the correlation function outside the strip $S$ exists or not, Eq.~(\ref{eq:8d}) holds for all Paley-Wiener functions $f$ with the property that the support of $\hat{f}$ is contained in the interval $(-2\tau_0, 2\tau_0)$. 
\end{3}

\section{A look at the numerical implementation}
In this section, we show that if the support of the distribution $\hat{f}(\kappa)$ is further restricted to be contained in the interval $(-\tau_0, \tau_0)$, then the generalized moment $\int_\mathbb{R} f(\omega)dP(\omega)$ can be evaluated exponentially fast from the short-time correlation function. To motivate the further restriction of the support, let us start by putting together Eqs.~(\ref{eq:4a}) and (\ref{eq:5a}) in a single formula
\begin{equation}
\label{eq:1f}
\int_\mathbb{R} f(\omega) dP(\omega) = \int_0^\infty e^{-s}\left[\frac{1}{2\pi i}\oint_\gamma f\left(\frac{s}{z}\right)\frac{C(-iz)}{z}dz\right]ds.
\end{equation}
We have explicitly utilized the square parenthesis in order to emphasize that the right-hand side must be regarded as an iterated integral, if the results from the preceding sections are to hold. The contour integral in the square parenthesis, that is the value of $\mathcal{P}_f(s)$, is calculated by quadrating an highly oscillatory integrand, a scenario that might lead to numerical difficulties. For fixed $s$, results of Fornaro\cite{For73} show that the quantity $\mathcal{P}_f(s)$ can be approximated exponentially fast by numerical quadrature. If we pick the contour $\gamma$ to be a circle  of radius $\tau < \tau_0$ centered about the origin, we  obtain
\begin{equation}
\label{eq:2f}
\mathcal{P}_f(s) = \int_0^1 f\left(s\,\tau^{-1}  e^{-i2\pi \phi}\right)C(-i\tau e^{i2\pi \phi})d\phi.
\end{equation}
According to Fornaro's Corollary~5.1 (the corollary will be stated below), the one-dimensional integral in Eq.~(\ref{eq:2f}) can be approximated by trapezoidal quadrature exponentially fast. Nevertheless, the rate of convergence is different for different values of $s$ and generally worsens in the limit $s \to \infty$. 

Not surprisingly, the negative scenario exposed in the preceding paragraph can be avoided provided that the Paley-Wiener functions are further restricted in such a way that the iterated integral appearing in Eq.~(\ref{eq:1f}) is also valid as a double integral (i.e., the conditions of the Fubini-Tonelli theorem are satisfied). In fact, in slightly more restrictive conditions, we show that Fornaro's corollary implies that the left-hand side of Eq.~(\ref{eq:1f}) can be approximated exponentially fast from the short-time values of the correlation function. Let us assume that $f(z)$ is a Paley-Wiener function obtained from a distribution $\hat{f}(\kappa)$ with support included in the interval $(-\tau_0, \tau_0)$. Thus, $B < \tau_0$ and we take $\tau$ such that $B < \tau < \tau_0$. From the Paley-Wiener theorem, we obtain that  
\[
|f(z^{-1}s)| \leq C (1+|z^{-1}s|)^N e^{B |\mathrm{Im}(z^{-1}s)|} \leq C(1+s/\tau)^N e^{B s/\tau}, \quad \forall \; |z| = \tau. 
\]
Since $B / \tau < 1$, it follows that $|f(z^{-1}s)|$ is absolutely integrable against $e^{-s}ds$. The Fubini-Tonelli theorem then tells us that
\begin{equation}
\label{eq:3f}
\int_\mathbb{R} f(\omega) dP(\omega) = \frac{1}{2\pi i}\oint_{|z|= \tau} \Phi_f\left(\frac{1}{z}\right) \frac{C(-iz)}{z}dz,
\end{equation}
where
\begin{equation}
\label{eq:4f}
\Phi_f(z) = \int_0^\infty e^{-s} f\left(zs\right)ds.
\end{equation}

Here, we will not be concerned with the numerical evaluation of $\Phi_f(z)$. This depends upon the choice of $f$ and, in most cases, the integral can be performed explicitly. As we have already mentioned in the second section, for orthogonal polynomials, the Gauss-Laguerre quadrature rule is more appropriate. The function $\Phi_f(z)$ is analytic in the strip defined by the equation $|\mathrm{Im}(z)| < 1/B$. The assertion follows from Proposition~\ref{Pr:1} by noticing that $\Phi_f(z)$ is a function of the type defined by Eq.~(\ref{eq:2d}), with the measure $dP(\omega)$ obtained from the density $e^{-\omega}$, if $\omega \geq 0$, and $0$, otherwise. For this measure, $\tau_0 = 1$, whereas the correlation (characteristic) function is given by the well known expression $C(t) = 1/(1-it)$. Eq.~(\ref{eq:6ds}) then gives an alternative way to evaluate $\Phi_f(z)$, namely
\begin{equation}
\label{eq:5f}
\Phi_f(z) = \int_{-B}^B  \frac{\hat{f}(\kappa)}{1-iz\kappa}d\kappa.
\end{equation}   

We now go back to Eq.~(\ref{eq:3f}) and replace the right-hand side integral with a trapezoidal quadrature sum. We obtain 
\begin{equation}
\label{eq:6f}
\int_\mathbb{R} f(\omega) dP(\omega) \approx \frac{1}{N+1}\sum_{k = 0}^N \Phi_f\left(\tau^{-1}e^{-i2\pi k /(N+1)}\right) C\left(-i\tau e^{i2\pi k /(N+1)}\right).
\end{equation}
In the notation of Fornaro, if we let $g(z) = f(1/z)C(-iz)/z$, then the difference between the left-hand side and the right-hand side of Eq.~(\ref{eq:6f}) is denoted by $E_N(g)$. Notice that the function $f(1/z)$ is analytic in the domain $|z| > B$, whereas the function $C(-iz)$ is analytic in the domain $|z| < \tau_0$. It follows that $g(z)$ is analytic on the annulus $B < |z| < \tau_0$. If $0 < \rho_1 < 1 < \rho_2$ are chosen such that $B < \rho_1 \tau < \tau < \rho_2 \tau < \tau_0$, then Corollary~5.1 of Ref.~\onlinecite{For73} asserts that 
\begin{equation}
\label{eq:7f}
E_N(g) = \frac{\tau}{2\pi i} \oint_{|z| = \rho_2} \frac{g(z\tau)dz}{1-z^{N+1}} + \frac{\tau}{2\pi i} \oint_{|z| = \rho_1} \frac{z^{N+1}g(z\tau)dz}{z^{N+1}-1}. 
\end{equation}
If $M_1, M_2 > 0$ are bounds of $g(z)$ on the circles of equations $|z| = \rho_1 \tau$ and $|z| = \rho_2\tau$, respectively, then Eq.~(\ref{eq:7f}) implies
\begin{equation}
\label{eq:8f}
|E_N(g)| \leq \frac{\tau M_2 \rho_2}{\rho_2^{\phantom{1}N+1}-1} + \frac{\tau M_1 \rho_1^{\phantom{1}N}}{1-\rho_1^{\phantom{1}N+1}},
\end{equation}
from which the exponential decay to zero follows quickly if one only remembers that $\rho_1 < 1$ and $\rho_2 > 1$. 

The result of Fornaro generalizes previous error estimates of Lyness and Delves.\cite{Lyn67} It can produce accurate estimates of the error, provided that the upper bound expressed by Eq.~(\ref{eq:8f}) is more carefully established. However, such error estimates are not within the scope of the present paper.

\section{Conclusions}
We have established that generalized moments of spectral distributions against Paley-Wiener type functions can be computed from short-time information via Eqs.~(\ref{eq:4a}) and (\ref{eq:5a}). However, it is required that the correlation function  be  defined by continuity on the entire complex place with the possible exception of two branch cuts on the imaginary axis. The values outside the small open disk about the origin do not have to be known, yet their existence is important. Our effort to prove Theorem~\ref{Th:2} for distributions was motivated by the fact that we wanted to maintain the polynomials and the complex exponentials of the form $\exp(izt)$ in the class of admissible functions. Indeed, these functions are of Paley-Wiener type because they are the Fourier-Laplace transform of Dirac's delta function or its distributional derivatives.

In practical applications, we will probably be more interested in the case when the distributions are represented by some integrable functions $\hat{f}(\kappa)$ vanishing outside some bounded interval $[-B, B]$. If the function is also infinitely differentiable on $[-B,B]$, say, if it is
\begin{equation}
\label{eq:1e}
\hat{f}(\kappa) = \left\{ \begin{array}{ll}
\exp\left[-\kappa^2 /(B^2-\kappa^2)\right], & \mathrm{if}\ |\kappa|<B,\\
0, & \mathrm{otherwise},
\end{array}\right.
\end{equation}
 the original version of the Paley-Wiener theorem tells us that, for each $N = 1, 2, \ldots$, there is a constant $C_N > 0$ such that
\[
|f(z)| \leq C_N (1+|z|)^{-N} e^{B|\mathrm{Im}(z)|}.
\]
This says that $f(z)$ decays faster than any polynomial on the real axis. Although, they cannot compete with a Gaussian in terms of decay of the tail, the Paley-Wiener functions obtained by transforming arbitrarily smooth and compactly supported functions are numerically more useful than  polynomials or  complex exponentials. Due to their integrability, they can form delta sequences and be utilized to generate representations of the type given by Eq.~(\ref{eq:6a}).  In this respect, it is useful to remember that, if $\sigma > 0$ and $\omega_0 \in \mathbb{R}$, then $f((z - \omega_0)/\sigma)$ is a Paley-Wiener function as well. This follows easily from the equality
\[
f((z - \omega_0)/\sigma) = \int_\mathbb{R} \hat{f}(\kappa) e^{i\kappa(z - \omega_0)/\sigma}d\kappa = \int_\mathbb{R} \left[ \sigma \hat{f}\left(\sigma\kappa\right)e^{-i\kappa\omega_0/\sigma}\right] e^{i\kappa z}d\kappa.
\]

Again, we mention that if the correlation function cannot be extended analytically beyond the strip $S$, the integral transformation is still valid provided that the support of the distribution $\hat{f}(\kappa)$ is included in the interval $(-2\tau_0, 2\tau_0)$. If the support of the distribution $\hat{f}(\kappa)$ is further restricted to be included in the interval $(-\tau_0, \tau_0)$, then the evaluation of the generalized moments $\int_\mathbb{R} f(\omega)dP(\omega)$ can be performed exponentially fast by means of Eq.~(\ref{eq:6f}). Although we lose the ability of forming delta sequences, we still have the possibility of using Paley-Wiener functions that decay to infinity. Such functions are more useful than polynomials in the context of maximum entropy techniques.\cite{Jar96, Kri99, Tag94}

 \begin{acknowledgments}
 This work was supported in part by the National Science Foundation Grant No. CHE-0345280 and by the Director, Office of Science, Office of Basic Energy Sciences, Chemical Sciences, Geosciences, and Biosciences Division, U.S. Department of Energy under Contract No. DE-AC02-05CH11231.
 \end{acknowledgments}

\appendix

\end{document}